\documentclass[journal=jacsat,manuscript=article]{achemso}
\usepackage{color}
\usepackage{xr}
\usepackage{amsmath,amsfonts,amssymb}
\usepackage{graphicx}
\newcommand*{\vect}[1]{\boldsymbol{#1}}

\usepackage[T1]{fontenc}
\usepackage{xr}

\author{Pavel I. Zhuravlev}
\affiliation{Biophysics Program, Institute for Physical Science and Technology, Department of Chemistry \& Biochemistry, University of Maryland, College Park, MD, 20742}
\author{Michael Hinczewski}
\affiliation{Department of Physics, Case Western Reserve University, Cleveland, OH, 44106}
\author{D.Thirumalai}
\affiliation{Department of Chemistry, The University of Texas, Austin, TX, 7712}

\title{Low force unfolding of a single-domain protein by parallel pathways}

\begin{document}
\maketitle
\begin{abstract}
Deviations from linearity in the dependence of the logarithm of protein unfolding rates, $\log k_u(f)$, as a function of mechanical force, $f$, measurable in single molecule experiments, can arise for many reasons.   In particular, upward curvature in $\log k_u(f)$ as a function of $f$ implies that the underlying energy landscape must be multidimensional with the possibility that unfolding ensues by parallel pathways. Here, simulations using the SOP-SC model of a wild type $\beta$-sandwich protein and several mutants, with immunoglobulin folds, show upward curvature in the unfolding kinetics. There are substantial changes in the structures of the transition state ensembles as force is increased, signaling a switch in the unfolding pathways.  Our results, when combined with previous theoretical and experimental studies, show that parallel unfolding of structurally unrelated single domain proteins can be determined from the dependence of $\log k_u(f)$ as a function of force (or $\log k_u[C]$ where $[C]$ is the denaturant concentration). 
\end{abstract}

\doublespacing
\section*{Introduction}
Since the pioneering studies on forced-unfolding of titin over twenty years ago \cite{Rief97Science,Kellermeyer97Science,Tskhovrebova97Nature}, single molecule pulling experiments have generated a wealth of data that continue to provide unprecedented glimpses into the response of biomolecules and their complexes to mechanical forces. In a typical pulling experiment \cite{Rief10PNAS,Cecconi05Science,Woodside08COCB,ZhuangCOSB03}, mechanical force ($f$) is applied to two points on a protein, and the response is monitored as a function of $f$. The experiments (optical tweezers or atomic force microscopy (AFM)) could be done at a constant $f$ (or similarly constant extension) or a constant loading rate, $r_f$.  The constant $f$ experiments generate a folding trajectory in which the protein (or RNA) could hop between a number of states accessible at $f$. If during the observation time, which can be extremely long, the system is ergodic then both the equilibrium and $f$-dependent rates between all the relevant states can be computed, provided some corrections due to the experimental setup are taken into account \cite{Hinczewski13PNAS,Ramm14PNAS}   In the constant loading rate mode, sometimes referred to as dynamic force spectroscopy, the distribution of rupture forces is measured by linearly increasing $f$ at a constant loading rate, $r_f = \frac{df}{dt}$. In many instances one encounters slip bonds in which plots of the most probable rupture force ($f^*$)  as a function of $\log r_f$ ($[\log r_f, f^*]$ plot) or the dependence of unfolding rate ($k_u(f)$) on $f$, (referred to as a $[\log k_u(f),f]$ plot) are linear, which is explained using the Evans-Ritchie (for constant $r_f$) theory \cite{Evans99BJ,Evans01ARBBS} or the Bell model (constant $f$) \cite{Bell78Science}. However, there are a number of examples (forced unfolding of monomeric proteins \cite{Jagannathan12PNAS,Zhuravlev16PNAS} and rupture of cell adhesion and other complexes \cite{Wojcikiewicz06Biomacro}), which show curvature or change in slope in the $[\log r_f, f^*]$ \cite{Merkel99Nature} and  $[\log k_u(f),f]$ plots. The data in these experiments could arise for the following reasons: (i) The location ($x^{\ddagger}$) of the transition state (TS) in the one dimensional free energy profiles as a function of extension, conjugate to $f$ and accessible in pulling experiments, moves with $f$ or $r_f$ due to an interplay between force and curvature of the energy landscape\cite{Hyeon03PNAS,Dudko06PRL,Dudko08PNAS}. (ii) The upward curvature is a consequence of unfolding occurring by parallel pathways in a strongly multidimensional landscape~\cite{Zhuravlev16PNAS}, which is surely the case in the SH3 domain~\cite{Jagannathan12PNAS}. (iii) An extremely interesting and counter intuitive scenario arises when the upward curvature is so pronounced that it gives rise to catch bonds, observed in a number of protein complexes \cite{Zhu03Nature,Chakrabarti17JSB}.  In this case as well as in (ii) the extension alone cannot serve as a reaction coordinate.  (iv) Rupture of protein complexes (biotin-streptavidin~\cite{Merkel99Nature,Hyeon12JCP}) involves overcoming two sequential barriers with two TS locations, one at low forces (large value of $x^{\ddagger}$)  and the other (small value of $x^{\ddagger}$) at high forces. (v) In forced-rupture of certain cell adhesion complexes, it is possible to construct a collective one dimensional coordinate~\cite{Toan18JCP}, which reflects the change in the slopes observed in the $[\log r_f, f^*]$ plots. These studies show that new physics  of biological molecules and complexes could be teased out using the data from single molecule experiments complemented by theory and well-designed simulations~\cite{Hinczewski16PNAS,Zhuravlev16PNAS}.

%Here, we revisit forced-unfolding of the single domain $\beta$-sheet protein immunoglobulin domain (I27) (Fig. \ref{fig:native}, in order to determine whether the unfolding occurs in a multidimensional landscape. I27 is a constituent of Titin, a giant protein found in the sarcomere, and  is composed of hundreds of domains, most of which have immunoglobulin type fold.  
%In addition, titin has a major biological function, which is the control of muscle elasticity. Therefore, the mechanical properties of titin, and those of immunoglobulin domain (I27) are important.  For instance, in muscular stress fibers, the mechanical force is applied to the ends of I27. 

%A pioneering series of single-molecule pulling experiments\cite{Rief:1997ve,Rico:2013fq} as well as molecular dynamics simulations at ultra-high forces and/or ultra-fast pulling rates\cite{Lee:2009fg} were used in the context of mechanics of titin and its major components, a variety of structurally similar immunoglobulin domains, such as I27. Titin, a giant protein found in the sarcomere, and  is composed of hundreds of domains, most of which have an immunoglobulin type fold.  In addition, titin has a major biological function, which is the control of muscle elasticity. Therefore, the mechanical properties of titin, and those of immunoglobulin domain (I27) in particular, are important. 
%For instance, in muscular stress fibers, mechanical force is applied to the ends of I27. 

\begin{figure}
	\includegraphics[width=0.9\textwidth]{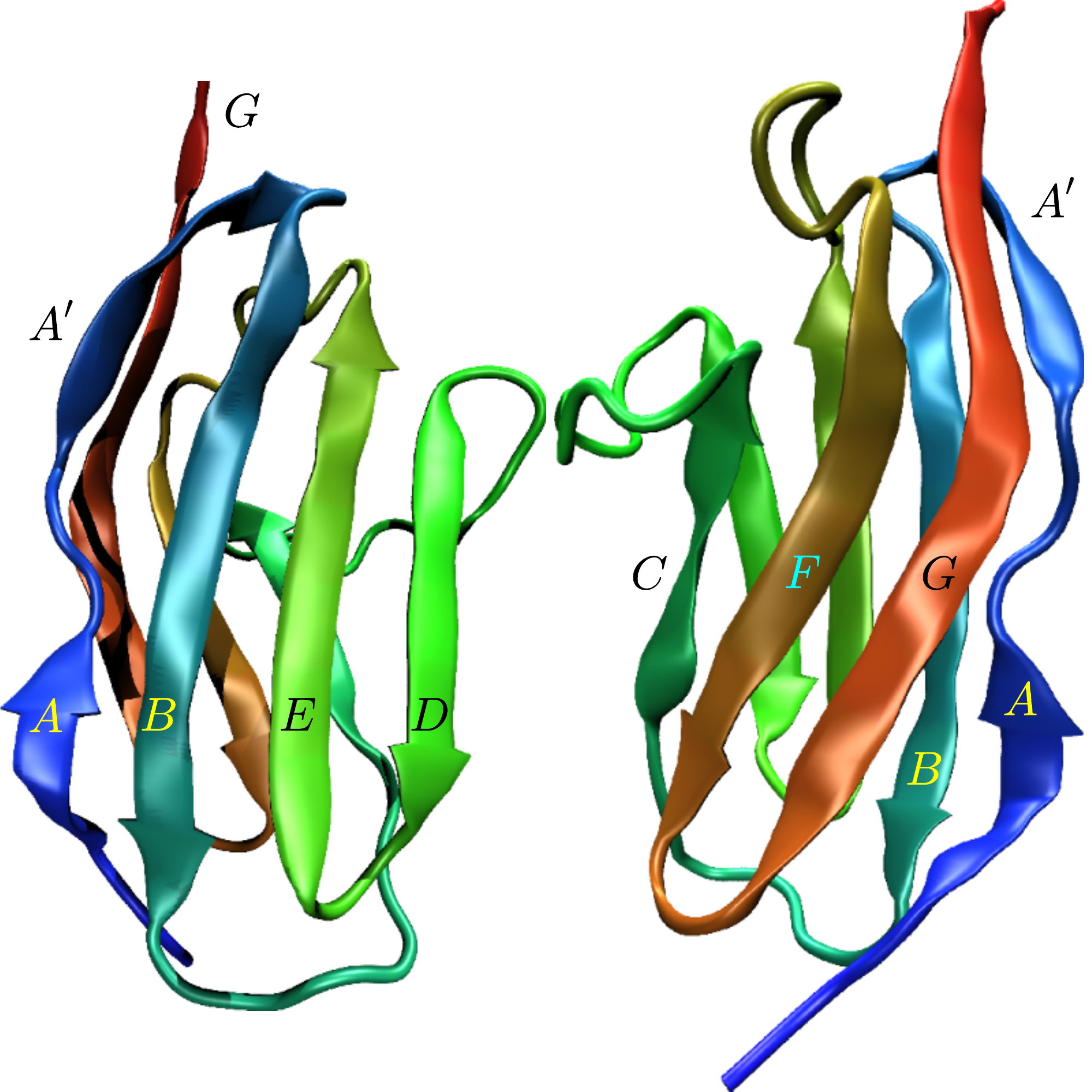}
	\caption{I27 is a 89 residue single domain protein consisting of two $\beta$-sheets $ABED$ and $CFGA^\prime$, made up of eight $\beta$-strands $A$, $A^\prime$, $B$, $C$, $D$, $E$, $F$ and $G$. Two orientations are shown for clarity.\label{fig:native}}
\end{figure}

Here, we revisit forced-unfolding of a single domain $\beta$-sheet immunoglobulin domain (I27) (Fig. \ref{fig:native}) in order to determine whether the low force unfolding occurs in a multidimensional landscape. It could be argued that the magnitudes of such high forces used in a majority of previous experimental and computational studies were not biologically relevant. However,  the use of high forces was warranted in order to demonstrate how such experiments can probe the unfolding of a single protein under mechanical forces. Due to its mechanical resilience, I27 unfolding is not easily realized in low-force single molecule spectroscopy, either in the laser optical trap (LOT) setup, or \textit{in silico} at room temperature in the absence of denaturants. Because $[\log k_u(f),f]$ plots for I27 were not readily available at low forces it has been difficult to determine whether forced-unfolding proceeds by parallel pathways with a switch from one to another at a critical force.

The earliest demonstration~\cite{Wright03NSB} that I27 unfolds by parallel pathways used the chemical denaturant guanidinium chloride (GdmCl) as a perturbation. In the wild type (WT) and several mutants, it was found that $\log k_u$ as a function of GdmCl concentration exhibited upward curvature. Using this data it was surmised that I27 must unfold by parallel pathways as GdmCl concentration is increased. Subsequently, other experiments on different proteins  have established the presence of upward curvature in the unfolding kinetics as the denaturant concentration is increased \cite{Aksel14BJ,Jha09PNAS, Aghera13Biochem,Zaidi97NSMB}.  

More recently, it has been shown using optical tweezer experiments that the $[\log k_u(f),f]$ plot for the SH3 domain exhibits upward curvature \cite{Jagannathan12PNAS}, which was interpreted in the same way as in GdmCl-induced unfolding of I27. That upward curvature in $[\log k_u(f),f]$ (and by analogy any perturbation that couples linearly to the protein conformations) implies that unfolding could occur by parallel pathways was rigorously demonstrated in general and for SH3 in particular \cite{Zhuravlev16PNAS}.% which in retrospect supports the interpretation provided previously \cite{Wright03NSB}.  %{\bf\color{red} Comment: like in the abstract comment, should we clarify that upward curvature in the absence of catch-bond-like behavior implies parallel pathways?  Because upward curvature by itself could still be seen for a catch-bond single pathway case.}

%Here, we revisit forced-unfolding of a single domain $\beta$-sheet immunoglobulin domain (I27) (Fig. \ref{fig:native}) in order to determine whether the low force unfolding occurs in a multidimensional landscape. The magnitudes of forces used in a majority of previous studies were not biologically relevant, but their use was warranted in order to demonstrate how such experiments can probe the unfolding of a single protein under mechanical forces. Due to its mechanical resilience, I27 unfolding is not accessible in low-force single molecule spectroscopy, either in the laser optical trap (LOT) setup, or \textit{in silico} at room temperature in the absence of denaturants. Because $[\log k_u(f),f]$ plots for I27 were not readily available at low forces it has been difficult to determine whether forced-unfolding proceeds by parallel pathways with a switch from one to another at a critical force.

In this paper, we report computer simulations of I27 unfolding at low forces (0 -- 50 pN) at elevated temperature, for the wild type and a series of single-point mutants.
We show that, at least for some of the mutants of I27 the $[\log k_u(f), f]$ plots show upward curvature, suggesting that the underlying energy landscape is strongly multidimensional, especially at low forces. Elsewhere \cite{Zhuravlev16PNAS} we showed that if $k_u(f)$ violates either $\frac{d}{df}\log k_f \ge 0$ or $\frac{d^2}{df^2}\log k_f \le 0$ then the underlying energy landscape is strongly multi-dimensional (SMD).   Forced unfolding in such a SMD landscape likely involves a switch in the pathway at a critical force (possibly over a small force range). Below the critical force, I27 unfolds predominantly along one route and at higher forces it follows a different pathway.   We analyze and compare transition state ensembles (TSE) at 5 pN and 35 pN, using the $P_{fold}$ analysis \cite{Du98JCP}, showing the change in the preferential order in which tertiary structure elements rupture. Thus, along with switch in the pathway there is a change in the TSEs. Taken together our study shows that careful theoretical and computational analysis of data from single molecule pulling experiments could be used to determine the dimensionality of the energy landscape of proteins provided $[\log k_u(f), f]$ is known for a broad range of $f$.

\section*{Results}

We performed molecular simulations using the  coarse-grained (SOP-SC) model \cite{Hyeon06Structure,Liu18JPCB} of titin I27 domain (PDB 1TIT) by applying a constant force to the ends of the protein. The goal of the simulations was to explore the force range, inaccessible to either AFM (which intrinsically uses high forces) or LOT setups (because unfolding times for I27 at low force are likely too long to be observed in LOT). 
The problem of very long unfolding times is particularly exacerbated in computer simulations, even using SOP-SC simulations.  To alleviate this problem, we performed simulations at 400 K.

\begin{figure}
	\includegraphics[width=0.9\textwidth]{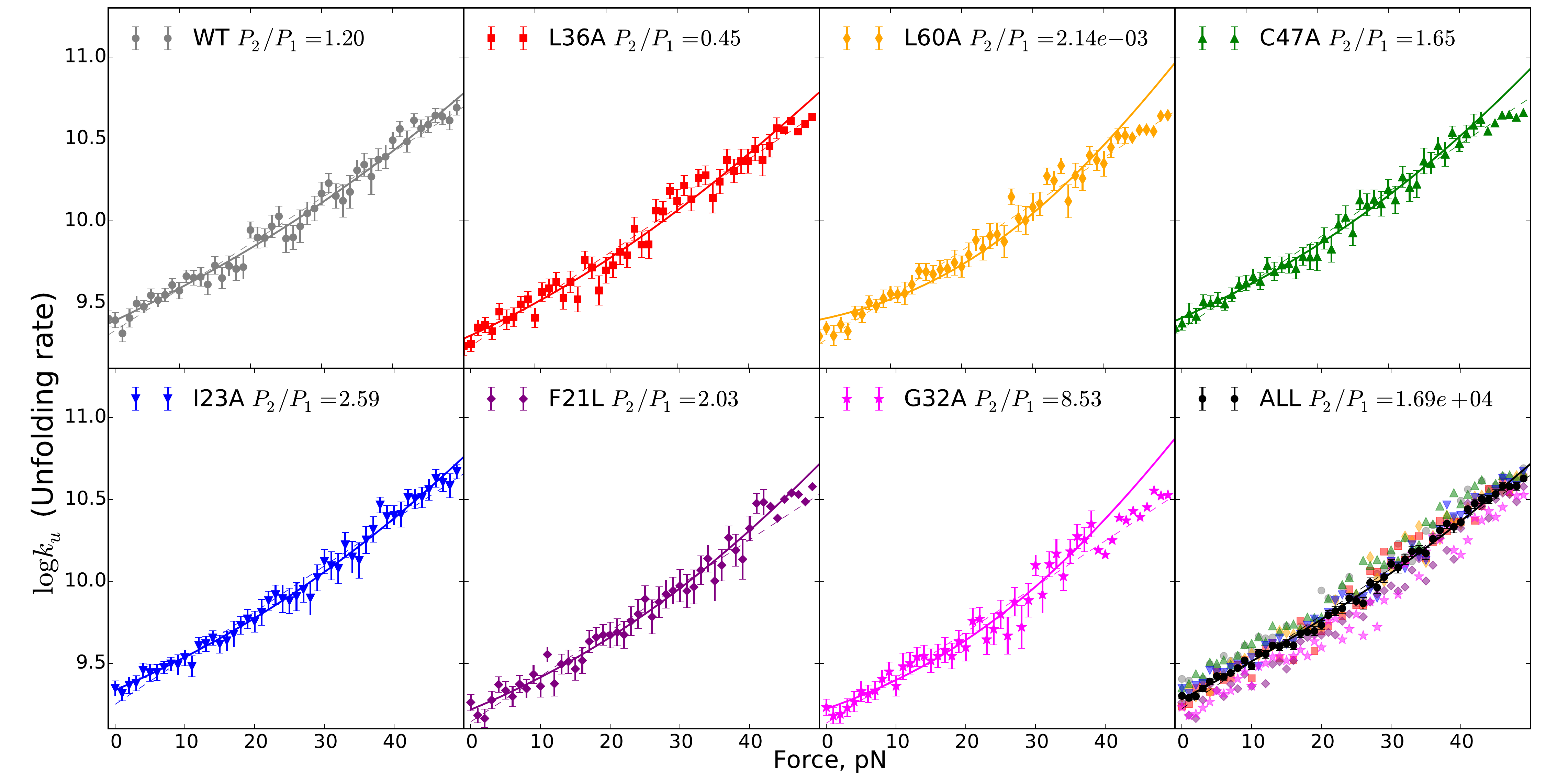}
	\caption{The unfolding rates of titin I27 domain and its mutants. Each point is calculated from the average unfolding time over 50 trajectories. For each mutant, the $[\log k_u(f),f]$ dependence is fitted with a single- (dashed lines) and a double-exponential (solid lines) model. The selection of fitting model is advised by the Akaike Information Criterion. The ratio of probabilities that particular model captures the data better than the other one ($P_2/P_1$ with $P_2$ ($P_1$) being the double (single) exponential fit) are given in the legends. Only L36A and L60A mutants data is better explained by single exponential fit. (Note: For L36A,G32A,L60A,C47A,F21L there is another change in slope around 35-40 pN, hence only the part of the dependence up to that point is fitted. The points that do not contribute to the shown fit and Akaike probability ratio calculations are shown without the error bars. However, using all the data (i.e. up to 50 pN) to fit these five mutants, results in single exponential model being selected (i.e. no curvature), leaving only WT and I23A with curvature.) The last panel uses all the data points, as if they were all for the same molecule, as a proxy (with the argument that changes to the force field for simulating different mutants are tiny) to assess the presence of upward curvature following from tertiary structure of titin I27 domain. With seven times more statistics, the Akaike model selection is strengthened and much more significant statistically, rather than disappeared, as evidenced by a four orders of magnitude higher $P_2/P_1$.\label{fig:8panel}}
\end{figure}

\textbf{Curvature in the $[\log k_u,f]$ plots:}
We generated 100 unfolding trajectories for the WT and each of the six mutants in order to calculate the  $[\log k_u(f),f]$ plots. As shown before \cite{Zhuravlev16PNAS}, the existence of upward curvature in these plots would be an unmistakable indication of strongly multidimensional unfolding dynamics.  One consequence is parallel rupture unfolding pathways. %I27 is a well studied domain which was shown to exhibit these rich unfolding dynamics in denaturant induced unfolding\cite{Wright:2003cs}.
Fig. \ref{fig:8panel} shows the $[\log k_u(f),f]$ plots for the wild-type and five of the mutants, from the set of mutants reported previously~\cite{Wright03NSB}.
When there is no curvature, the $[\log k_u(f),f]$ is linear, and can be fit with the single-exponential Bell model, $k_u(f)=k_0\exp\left(\frac{fx}{k_BT}\right)]$.
To identify the presence of upward curvature in the plot, we compare the Bell model fit to one with a double exponential $k_u(f)=k_1\exp\left(\frac{fx_1}{k_BT}\right)+k_2\exp\left(\frac{fx_2}{k_BT}\right)$ using the Akaike Information Criterion (AIC) \cite{Burnham02Book}.

AIC is an estimator of relative quality of statistical models for a given set of data, based on information theory \cite{Burnham02Book}. It provides a means for model selection by estimating how much information (from the data) is lost when using the model to represent the process which produced those data. The AIC value of a model is $AIC=2k - 2\ln\hat{L}$, where $k$ is the number of parameters in the model and $\hat{L}$ is the maximum value of the likelihood function of the model. The model that has the minimal $AIC$ is most likely to lose the least information from the data, and thus is selected. If the models have values $AIC_1, AIC_2, ...$, and we define $AIC_\mathrm{min}$ as the minimum of those values, then the quantity $\exp{(AIC_\mathrm{min} - AIC_i)/2}$ can be interpreted as proportional to the probability that the $i$-th model minimizes the estimated information loss. Then $P_2/P_1 = \frac{\exp{(AIC_\mathrm{min} - AIC_2)/2}}{\exp{(AIC_\mathrm{min} - AIC_1)/2}} =\exp{(AIC_1 - AIC_2)/2}$, for example, is the ratio of the probability $P_2$ that model 2 minimizes the information loss to the probability $P_1$ that model 1 minimizes the information loss. Applied to our case of choosing between the single- and double-exponential fits of $[\log k_u(f),f]$, we calculate the $P_2 / P_1$ ratio of probabilities that a double- or single- exponential model respectively minimizes the information loss.

If $P_2/P_1 > 1$, the double exponential model is more likely to minimize the information loss. If, in addition $x_2>x_1$, we conclude that the upward curvature fit is superior, implying strong dimensionality of the unfolding process.
(We avoid the temptation to over interpret the fitting parameters and use this procedure of fitting to a double-exponential model as a statistical means to establish the presence of the curvature.  Double exponential, which comports with the physics of force-induced unfolding, is one of the simplest models that exhibits upward curvature in $[\log k_u(f),f]$).  %{\bf\color{red} Comment: Playing devil's advocate, if we converted $P_2/P_1$ to a rough $p$-value by treating the single exponential case as the null model, then a $P_2/P_1$ ratio of 0.95/0.05 = 19 would be the conventional cutoff for ``statistical significance''.  Are we at risk with reviewers claiming none of our $P_2/P_1$ ratios reach this level (except the ALL case, discussed in the comment below)?}
%The statistical analysis of the data described above shows upward curvature in WT, I23A and F21L constructs.

At very high forces, we expect a linear $\log k_u(f)$ dependence, with the value of the slope that can be higher, lower or the same as the slopes we observe at  lower forces. It is not known where this switch to the linear regime should occur for any of the constructs.
In the case when the switch occurs below 50 pN (the upper limit of the curves in Fig. \ref{fig:8panel}), the double exponential model fit only makes sense up to the point of the switching. The data points not contributing to the fits in Fig. \ref{fig:8panel} are shown without error bars.
For L36A, G32A, L60A, C47A and F21L we used the fit up to the point (38-45 pN), visually identified as the change in slope.
WT and I23A have a curvature in the $[\log k_u(f),f]$ plots even when the fitting is extended to 50 pN, according to the AIC statistic,  which we use for selection of the fitting model (single vs. double exponential).
C47A, F21L and G23A have curvature when fitted up to $\sim$ 40 pN, where there is an apparent downward change in the slope.
L36A and L60A do not exhibit statistically significant curvature.

Since the changes in the force field for point mutations are small, it is tempting to test for the curvature treating all of the simulations as the simulations of the same molecule, thus increasing the available statistics. The bottom right panel of Fig.\ref{fig:8panel} shows the points for all the mutants on the same plot, and the black symbols represent the combined statistics. Akaike Information Criterion selects the model with curvature with very high probability, confirming the existence of parallel pathways in I27 unfolding. 
%\dcomment{ PZ and MH - we decided to comment on the rightmost panel to make our argument convincing but I forgot what that was..Could one of you fill it in, if needed}
%\pcomment{I don't think we wanted to add anything here? We just wanted to convince oursevles and decide whether to include it. There is also more information under the figure, which I think I added last month, not 5 years ago.}

%{\bf PZ and MH - to keep or not to keep?}
%In L36A the switch can be seen visually (at around 36 pN), and the dependence shows upward curvature if taken up to 36 pN, but no curvature (i.e. single exponential is more probable), when fitted up to 50 pN.
%The G32A mutant has the same dependence with the switch at 39 pN.
%L60A mutant does not show upward curvature even when considering dependence up to 35 pN.
%{\bf\color{red} Comment: but the mutants are physically different, so it seems a little strange to lump them all together.  For example, imagine if all the mutants showed single-exponential behavior, but with somewhat different exponents.  Then analyzing all the data at once would spuriously indicate multi-exponential behavior.  So I wouldn't necessarily trust that the high $P_2/P_1$ value for all the data is strong proof one way or another.  What do the experiments show for these mutants?}

\textbf{Secondary and tertiary structure elements:} 
I27 consists of eight $\beta$-strands (named $A$,$A^\prime$,$B$,$C$,$D$,$E$,$F$ and $G$ along the sequence).
They form two $\beta$-sheets: $ABED$ and $CFGA^\prime$ (Fig.\ref{fig:native}).
Rupturing of pairs of strands can be therefore characterized by parameters $\chi_{AB}$,$\chi_{BE}$,$\chi_{ED}$,$\chi_{CF}$,$\chi_{FG}$ and $\chi_{GA^\prime}$, where $\chi_{X Y}$ is the structural overlap for two parts of the protein $X$ and $Y$:  the fraction of broken native contacts between $X$ and $Y$\cite{Guo95BP},
\begin{equation}
	\chi_{XY}(\{\vect{r}\})=\frac{1}{M_{XY}}\sum_{\substack{i \in X\\ j \in Y}}\Theta\left(\left||\vect{r}_i-\vect{r}_j|-|\vect{r}_i^0-\vect{r}_j^0|\right|-\Delta\right).
\end{equation}
In the above equation, the summation is over the coarse-grained beads belonging to the parts $X$ and $Y$, $M_{XY}$ is the number of contacts between $X$ and $Y$ in the native state, $\Theta(x)$ is the Heaviside function, $\Delta=2\mathring{\mathrm{A}}$ is the tolerance in the definition of a contact, and $\vect{r}_{i,j}$ and $\vect{r}_{i,j}^0$, respectively, are the coordinates of the beads in a given conformation $\{\vect{r}\}$ and the native state.
To compare low-force and high-force unfolding pathways we analyzed the trajectories at 5 pN and 35 pN for the I23A mutant, which does exhibit upward curvature in the $[\log k_u(f),f]$ plot.

\textbf{Order of rupturing of $\beta$-strand pairs:}
Table \ref{table:fraction100} shows the percentage of trajectories in which one pair of $\beta$-strands (row) fully ruptured ($\chi$=1) before the other (column). 
The comparisons where there is a significant change between low and high force are marked in bold.
Most notably, at low force the $CF$ pair never breaks after $FG$ and almost never after $GA^\prime$.
At high force, $CF$ breaks after $FG$ and after $GA^\prime$ more frequently.
Also $CF$ almost always breaks after $AB$ at high force, but less frequently at low force.
The fraction of trajectories where $AB$ breaks after $ED$ changes from two thirds at low force to one third at high force.
With high probability, $BE$ strand rupture precedes $GA^\prime$ breakage at low forces, but in a third of the trajectories  $GA^\prime$ ruptures before $BE$ at high forces.
These observations are intuitively consistent with force being applied to the ends of the protein, $A$ and $G$ strands, making unfolding pathways starting with rupturing of pairs involving these two strands more accessible at high forces. In other words, high force promotes rupturing of $AB$ and $GA^\prime$ before other elements.

\begin{figure}\centering
	\includegraphics[width=1.2\textwidth]{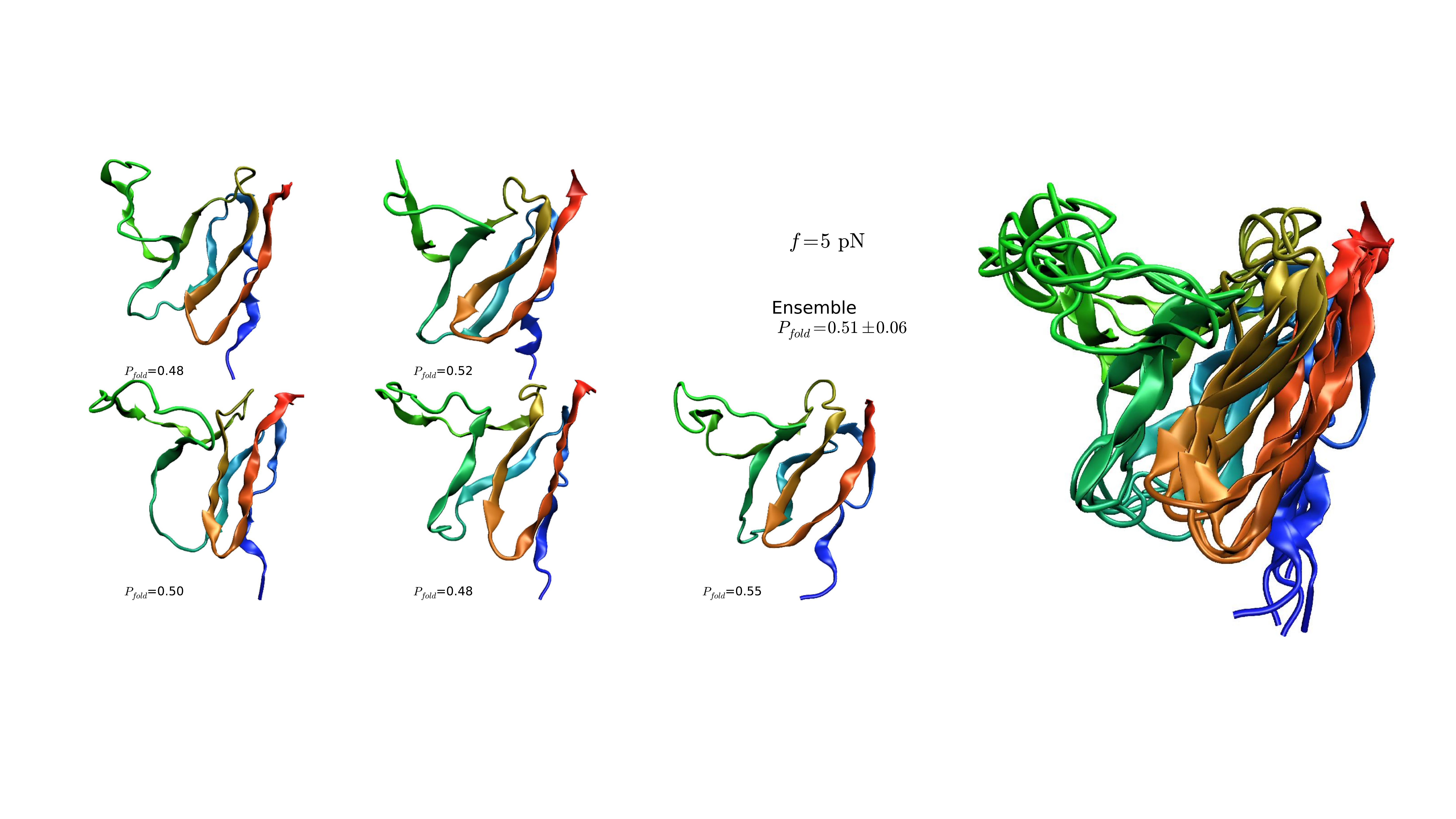}
	\caption{Representative TS structures and the calculated $P_\mathrm{fold}$ (below each structure) at 5 pN. The calculated value for the TSE is $P_\mathrm{fold}=0.51 \pm 0.06$.  $P_\mathrm{fold}=0.51 \pm 0.05$. Superposition of structures in the transition state ensemble. \label{fig:5align}}
\end{figure}

\begin{figure}\centering
	\includegraphics[width=1.1\textwidth]{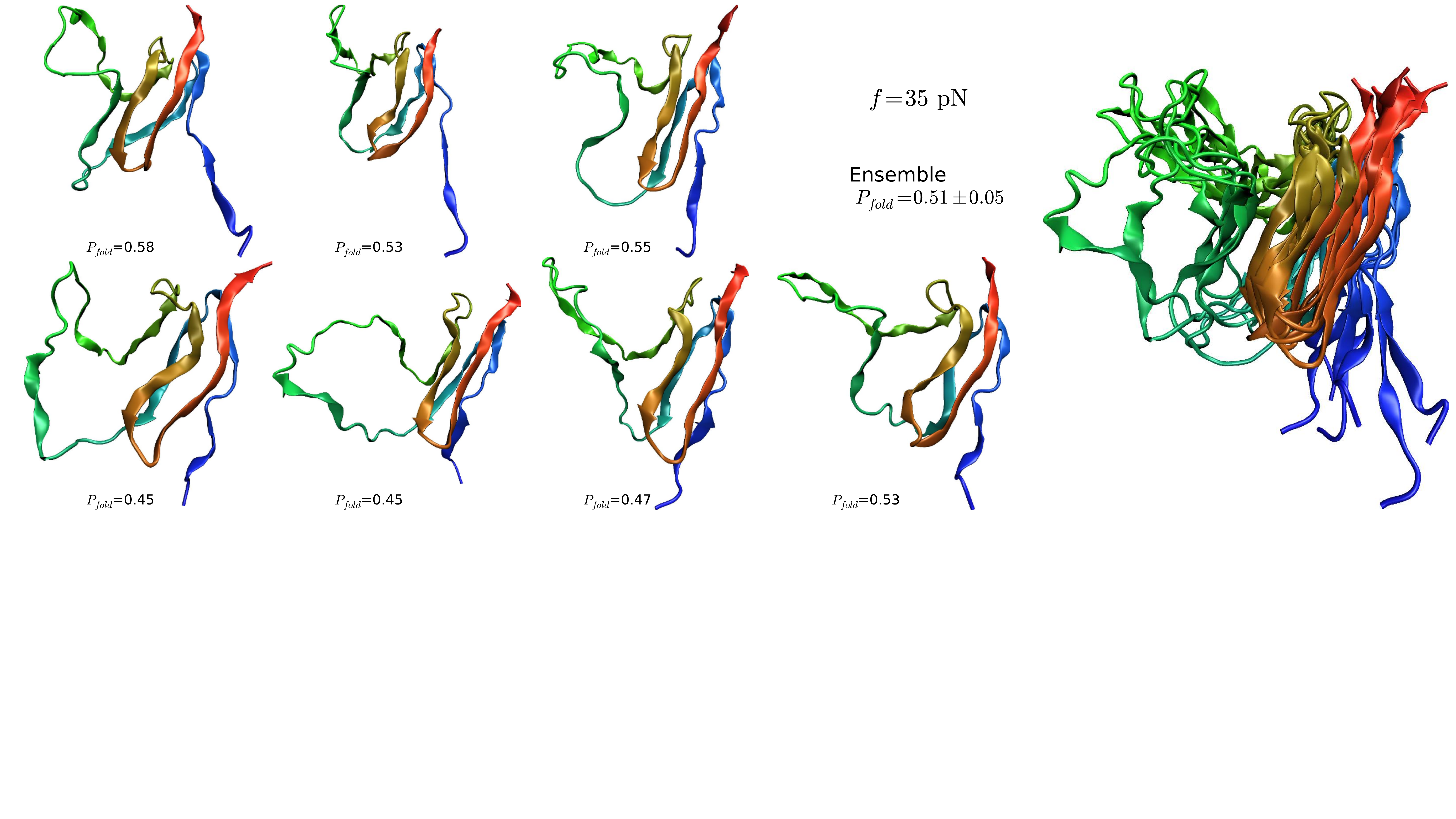}
	\caption{Same as Fig.\ref{fig:5align} except the value of the force is 35 pN. The average of $P_\mathrm{fold}=0.51 \pm 0.05$\label{fig:35align}}
\end{figure}

\begin{figure}
\includegraphics[width=1.1\textwidth]{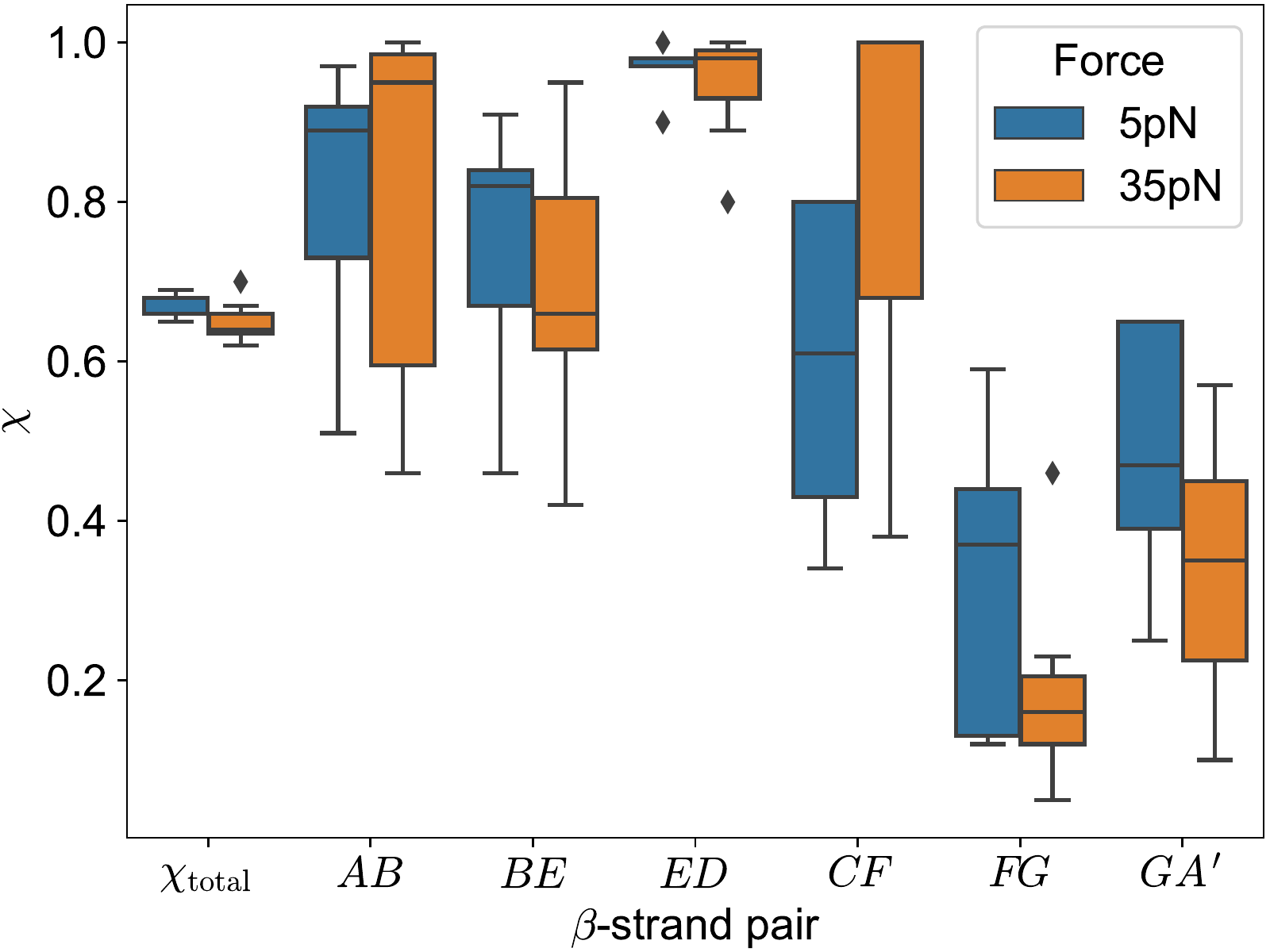}
\caption{The comparison of ruptured tertiary structure in TSE at 5 and 35 pN (box plot)\label{fig:boxchi}}
\end{figure}

\textbf{Transition state ensembles (TSEs) at high and low force:}
Another way to look at the unfolding pathway is to identify the transition state ensemble (TSE).
We define TSE through $P_{fold}$ \cite{Du98JCP}, which is the probability to reach the folded state before the unfolded state, starting from a given conformation. If $P_{fold}=0.5$ for some conformation, then this conformation belongs to the TSE.
We select TSE candidates from the saddle point of the $(\chi, E)$ free energy surface, where $E$ is energy and $\chi$ is the full overlap parameter ($\chi_{XY}$ where both $X$ and $Y$ correspond to the whole protein)
We obtain the free energy surface by calculating the histograms of the unfolding trajectories at 5 and 35 pN.
Running 10 trajectories for each of the candidates, we select a smaller set of candidates where $0.3 \leq P_{fold} \leq 0.7$, and then run 50 more trajectories for each candidate in the smaller set.
We obtain the TSE as the ensemble of candidates with $0.4 < P_{fold} < 0.6$ calculated from the 60 trajectories for each structure.
The TSEs for low and high force unfolding (TSE-L and TSE-H) are shown in Fig. \ref{fig:5align} (5 pN) and Fig. \ref{fig:35align} (35 pN).
Visually analyzing the structures, one can see $CF$ to be more disrupted in TSE-H.
Also, there are structures in TSE-H with $AB$ fully ruptured, but none in TSE-L.
To quantify these observations (along with the same observations about other structural elements) we calculated pairwise $\chi$ for each of the TSEs (Tables \ref{table:fraction100}, \ref{table:fraction95}, and Fig. \ref{fig:boxchi}).
From the data, one can conclude that $CF$ and $AB$ are less structured in TSE-H (and there are structures in the ensemble where one of them is fully ruptured, but not both) than TSE-L.
On the other hand, $FG$ and $GA^\prime$ are more structured in TSE-H than in TSE-L.
These observations indicate that force induced unfolding of I27 is heterogeneous, rather than having a dominant pathway, and the details of the heterogeneity (e.g. contribution of different pathways) changes with the applied force.

%\section*{Discussion}

\section*{Conclusion}
We have shown that at low forces, accessible in LOT experiments, the unfolding of I27 does not occur over a single pathway, but is heterogeneous. The finding that low forced unfolding must occur by parallel pathways is based on the the observation of upward curvature in the $[\log k_u(f), f]$ plot and by establishing that TSE structures, determined without assuming an underlying reaction coordinate, change as $f$ is increased. 
Our prediction can be confirmed experimentally, by measuring the $f$-dependence of $\log k_u(f)$ plot, like we did in simulations.

Our simulations provide structural insights into heterogeneities and differences of unfolding pathways at low and high forces.
We predict that single molecule spectroscopy of the mechanically resilient titin I27 domain (both wild types and mutants) at low forces and elevated temperature should provide  evidence that the energy landscape is strongly multidimensional. In addition, we show that there ought to be a switch in the folding pathway and associated changes in the structures of the transition state ensemble. Because it has not been done, we assume that varying temperature by a significant amount in optical tweezer experiments must be difficult. This could be mitigated by performing pulling experiments in the presence of small amount of a mild denaturant (urea) as has been done for the SH3 domain \cite{Guinn15NatComm}.

\section*{Methods}
We used the SOP-SC model\cite{Liu11PNAS,Zhuravlev14JMB}, which  is a coarse-grained model for proteins with two interaction beads per residue to simulate force-induced unfolding of I27. 
The interactions are based on the native structure, but the amino acid identity of each residue is also taken into account.
A mutation is applied to the PDB structure of the wild type, so the interactions are sorted into native and non-native subsequently. 
Thus, the set of native contacts remains the same, but a few interaction parameters change because of mutations. The interactions between the side chains are given by the statistical potential~\cite{Betancourt99ProtSci}. The details of the model and the simulations may be found elsewhere~\cite{Zhuravlev14JMB}.

We used maximum likelihood fitting to discriminate between a double exponential and single exponential (Bell) models for the dependence of the unfolding rate $k_u(f)$ on force. 
We compared the model fits by the Akaike Information Criterion \cite{Burnham02Book} in order to find out whether there is upward curvature in the $[\log k_u, f]$ plot, which would establish the multidimensionality of energy landscape leading to parallel unfolding pathways.

{\bf Acknowledgments:} This work was done while the authors were in the Institute for Physical Sciences and Technology at the University of Maryland.  This work was supported by the NSF CAREER MCB-BIO (1651560), the NIH (GM - 107703), the Welch Foundation Grant F-0019 through the Collie-Welch chair.

%\bibliographystyle{pnas}
%\bibliography{mybib3}
%\bibliography{all}

\clearpage
\begin{table}[htb!]
	\caption{Percentage of trajectories where the row contacts fully break after column contacts, at 5pN; at 35 pN}
	\label{table:fraction100}
	\centering
	\begin{tabular}{c|c|c|c|c|c|c}
	 &  $\chi_{AB}$  &  $\chi_{BE}$  &  $\chi_{ED}$  &  $\chi_{CF}$  &  $\chi_{FG}$  &  $\chi_{GA^\prime}$ \\ 
	\hline
	$\chi_{AB}$   &  &  14;  4 &  \textbf{66; 34} & \textbf{ 24;  4} &   0;  0 &   2;  2 \\ 
	\hline
	$\chi_{BE}$   &  86; 96 &  &  98; 94 &  62; 56 &   2;  8 &   \textbf{4; 30} \\ 
	\hline
	$\chi_{ED}$   &  34; 66 &   2;  6 &  &  12; 14 &   0;  2 &   2;  8 \\ 
	\hline
	$\chi_{CF}$   &  76; 96 &  38; 44 &  88; 86 &  &   \textbf{0;  8} &   \textbf{6; 22} \\ 
	\hline
	$\chi_{FG}$   &  100; 100 &  97; 91 &  100; 97 &  100; 91 &  &  71; 79 \\ 
	\hline
	$\chi_{GA^\prime}$   &  97; 98 &  95; 68 &  97; 92 &  93; 78 &  28; 20 &  \\
	\end{tabular}
\end{table}

\begin{table}[htb!]
	\caption{Percentage of trajectories where the row contacts break 95\% after column contacts, at 5pN; at 35 pN}
	\label{table:fraction95}
	\centering
	\begin{tabular}{c|c|c|c|c|c|c}
	 &  $\chi_{AB}$  &  $\chi_{BE}$  &  $\chi_{ED}$  &  $\chi_{CF}$  &  $\chi_{FG}$  &  $\chi_{GA^\prime}$ \\ 
	 \hline
	 $\chi_{AB}$   &  &  16;  4 &  80; 42 &  20;  6 &   0;  0 &   4;  4 \\ 
	 \hline
	 $\chi_{BE}$   &  84; 96 &  &  100; 96 &  44; 52 &   0; 16 &   6; 28 \\ 
	 \hline
	 $\chi_{ED}$   &  20; 57 &   0;  4 &  &   2; 10 &   0;  0 &   4;  4 \\ 
	 \hline
	 $\chi_{CF}$   &  80; 94 &  56; 48 &  98; 90 &  &   0;  8 &   4; 24 \\ 
	 \hline
	 $\chi_{FG}$   &  100; 100 &  100; 83 &  100; 100 &  100; 91 &  &  56; 73 \\ 
	 \hline
	 $\chi_{GA^\prime}$   &  96; 96 &  94; 72 &  96; 96 &  96; 74 &  44; 26 &  \\
	\end{tabular}
\end{table}

\clearpage

\clearpage
% \begin{figure}
% 	\includegraphics[width=0.9\textwidth]{separate}
% 	\caption{The unfolding rates of titin I27 domain and its mutants. Each point is calculated from the average unfolding time over 50 trajectories. For each mutant, the $\log k_u(f),f]$ dependence is fitted with a single- (dashed lines) and a double-exponential (solid lines) model, which are then compared by Akaike Information Criterion. The ratio of probabilities that particular model explains the data ($P_2/P_1$) are given in the legends. For L36A and G32A there is another change in slope around 35-40 pN, hence only the part of the dependence up to 36 (for L36A) and 39 (for G32A) pN is fitted. The fitting of the data for L36A and G32A up to 50 pN shows the single exponential model to be more probable (i.e. no curvature). \label{fig:separate}}
% \end{figure}
%
% \clearpage

% \begin{figure}
% 	\includegraphics[width=0.9\textwidth]{joint}
% 	\caption{The same plots as in Fig.\ref{fig:separate} on the same plot, for comparsion\label{fig:joint}}
% \end{figure}

%{\bf PZ - This figure confuses me. (1) Is it the case $P_2$ and $P_1$ are the actual probabilities and the Akaike score? (2) Why does $P_2/P_1$ greater than unity imply the the two exponential model is better? If this is correct then only L36A and L60A should be single exponential. (3) The last box labelled ALL - what does that mean and why is $P_2/P_1$ so large? Of course, I would be super surprised if you remember the answers!}
\newpage

%chiABchiED.png
%chiCFchiAB.png
%chiCFchiFG.png
%chiCFchiGap.png

% \begin{figure}
% 	\includegraphics[width=0.9\textwidth]{chiABchiBE.png}
% 	\caption{Difference in order of tertiary structure rupturing at 5 and 45 pN}
% \end{figure}
% \begin{figure}
% 	\includegraphics[width=0.9\textwidth]{chiABchiED.png}
% 	\caption{Difference in order of tertiary structure rupturing at 5 and 45 pN}
% \end{figure}
% \begin{figure}
% 	\includegraphics[width=0.9\textwidth]{chiCFchiAB.png}
% 	\caption{Difference in order of tertiary structure rupturing at 5 and 45 pN}
% \end{figure}
% \begin{figure}
% 	\includegraphics[width=0.9\textwidth]{chiCFchiFG.png}
% 	\caption{Difference in order of tertiary structure rupturing at 5 and 45 pN}
% \end{figure}
% \begin{figure}
% 	\includegraphics[width=0.9\textwidth]{chiCFchiGap.png}
% 	\caption{Difference in order of tertiary structure rupturing at 5 and 45 pN}
% \end{figure}
\clearpage
\clearpage
%\begin{figure}
%	\includegraphics[width=0.9\textwidth]{snapshots05.pdf}
%	\caption{Transition state ensemble at 5 pN. $P_\mathrm{fold}=0.51 \pm 0.06$\label{fig:5snapshots}}
%\end{figure}
%\begin{figure}
%	\includegraphics[width=0.9\textwidth]{snapshots35.pdf}
%	\caption{Transition state ensemble at 35 pN. $P_\mathrm{fold}=0.50 \pm 0.04$\label{fig:35snapshots}}
%\end{figure}

%{\bf PZ Could you combine the following two figures? Also what are the black dots?}
\clearpage
%\clearpage
%\begin{figure}
%\input{chistsebox.tex}
%\caption{The comparison of ruptured tertiary structure in TSE at 5 and 35 pN (box plot)\label{fig:boxchi}}
%\end{figure}
%\clearpage
%\begin{figure}
%\input{chistseviol.tex}
%\caption{The comparison of ruptured tertiary structure in TSE at 5 and 35 pN (violin plot)\label{fig:violinchi}}
%\end{figure}
	
\end{document}